\documentclass[pra,aps,twocolumn,floatfix,superscriptaddress,showkeys]{revtex4}  
\usepackage{graphicx}
\usepackage[ansinew]{inputenc}
\usepackage{array}
\usepackage{color}
\usepackage{amsmath}
\usepackage{amsxtra}
\usepackage{amstext}
\usepackage{amssymb}
\usepackage{latexsym}
\usepackage{dsfont}
\usepackage{verbatim}
\usepackage{comment}
\usepackage{epstopdf}
\usepackage{gensymb}

\begin{document}

\title{Enhanced Second Harmonic Generation from Coupled Asymmetric Plasmonic Metal Nanostructures}

\author{Bilge Can Yildiz}
\affiliation{Department of Physics, Middle East Technical University, 06800, Ankara, Turkey}
\affiliation{The Center for Solar Energy Research and Applications (GUNAM), Middle East Technical University, 06800, Ankara, Turkey}
\affiliation{Physics Unit, Atilim University, 06836 Ankara, Turkey}
\affiliation{Institute of Nuclear Sciences, Hacettepe University, 06800, Ankara, Turkey}

\author{Mehmet Emre Tasgin}
\affiliation{Institute of Nuclear Sciences, Hacettepe University, 06800, Ankara, Turkey}

\author{Musa Kurtulus Abak}
\affiliation{The Center for Solar Energy Research and Applications (GUNAM), Middle East Technical University, 06800, Ankara, Turkey}
\affiliation{Micro and Nanotechnology Program of Graduate School of Natural and Applied Sciences, Middle East Technical University, Ankara 06800, Turkey}

\author{Sahin Coskun}
\affiliation{The Center for Solar Energy Research and Applications (GUNAM), Middle East Technical University, 06800, Ankara, Turkey}
\affiliation{Department of Metallurgical and Materials Engineering, Middle East Technical University, 06800 Ankara, Turkey}

\author{Husnu Emrah Unalan}
\affiliation{The Center for Solar Energy Research and Applications (GUNAM), Middle East Technical University, 06800, Ankara, Turkey}
\affiliation{Micro and Nanotechnology Program of Graduate School of Natural and Applied Sciences, Middle East Technical University, Ankara 06800, Turkey}
\affiliation{Department of Metallurgical and Materials Engineering, Middle East Technical University, 06800 Ankara, Turkey}

\author{Alpan Bek}
\affiliation{Department of Physics, Middle East Technical University, 06800, Ankara, Turkey}
\affiliation{The Center for Solar Energy Research and Applications (GUNAM), Middle East Technical University, 06800, Ankara, Turkey}
\affiliation{Micro and Nanotechnology Program of Graduate School of Natural and Applied Sciences, Middle East Technical University, Ankara 06800, Turkey}

\date{\today}

\begin{abstract}
We experimentally demonstrate that two coupled metal nanostructures (MNSs), a silver nanowire and bipyramid, can produce $\sim$30 times enhanced second harmonic generation compared to the particles alone. We develop a simple theoretical model presenting the path interference effects in the nonlinear response of coupled MNSs. We show that the reason for such an enhancement can be the occurrence of a Fano resonance due to the coupling of the converter MNS to the long-lived mode of the attached MNS.   
\end{abstract}


\maketitle

\section{Introduction}

Typically, nonlinear optical effects appear at very high intensities, that is,  nonlinear optical properties of material surfaces can only be revealed by pulsed lasers.  In recent studies \cite{luk2010fano,Turkpence14,Salakhutdinov14}, material surfaces or interfaces decorated with metal nanostructures are suggested to act as nonlinear conversion agents since they enhance the local field amplitudes by several orders of magnitude. This property is expected to enable these efficient nonlinear converters to be utilized in many application areas, such as solar energy, molecular switching, photocatalysis, imaging, etc. 

Metal nanostructures (MNSs) exhibit absorption resonances at the optical frequencies. Such excitations may create hot-spots on the surfaces of MNSs. When a quantum dot (QD) is placed in the vicinity of these hot-spots, localized surface plasmons (LSPs) interact strongly with the attached QD. Presence of the QD introduces two absorption paths both lying in the plasmon spectral width \cite{Alzar02}, hence making them unresolvable. The two paths operate out of phase --one absorbing while the second emitting-- and result a dip in the absorption spectrum, where an absorption peak would be observed instead. Such transparencies in the plasmonic response are referred as Fano resonances \cite{Miroshnichenko10}. These are responsible for increased fluorescence of molecules \cite{Ayala-Orozco14} and increased lifetime of plasmonic oscillations \cite{Tasgin13} which makes coherent plasmon emission possible \cite{Noginov09,Stockman11}. Quantum coherence effects can be used to increase the sensitivity of ultrafast response nanolasers \cite{Voronine14} and can enhance the amplification \cite{Dorfman13,Voronine15}. 

Plasmon oscillations in MNSs concentrate the incident light to nm dimensions which yield strong enhancement in the field intensities \cite{Stockman11}. The enhancement in the intensity leads to the appearance of optical nonlinearities \cite{Zayats12} such as as enhanced Raman scattering \cite{Sharma12}, four wave mixing \cite{Genevet10,Renger11}, two photon absorption \cite{Cox13,Anton13,Racknor14}, and second harmonic generation (SHG) \cite{Thyagarajan13,Berthelot12,Wunderlich13,Gao11,Singh13,Walsh13}. Fano resonances with attached quantum objects can be utilized to enhance \cite{Thyagarajan13} and suppress \cite{Turkpence14} SHG process in plasmonic particles. The underlying mechanism relies on the cancellation of nonresonant frequency terms degrading the frequency conversion by hybridized paths \cite{Turkpence14}.

Fano-like resonances can also take place in two coupled classical oscillators \cite{Alzar02,Tassin12}, without quantum nature. It is also experimentally demonstrated that coupling with dark modes (which have longer lifetimes) \cite{panaro2014dark,Artar11,Cetin11} can result in Fano resonances.

In this paper, we investigate the second harmonic (SH) signal from coupled silver nanowire (AgNW) and silver bipyramid nanoparticle (AgNP). We observe that SH signal from the coupled system is highly enhanced compared to MNSs alone. We introduce a very simple theoretical model and show that such a factor of enhancement can be obtained via Fano resonances in the nonlinear response. We show that coupling the converter (AgNW) to the long-lived (10 times), e.g. dark \cite{panaro2014dark}, mode of an AgNP can enhance the SH signal in the similar amounts observed in our experiment. Interacting systems of MNSs are experimentally more controllable  \cite{Alzar02,Berthelot12,Tassin12,Artar11,Cetin11} compared to Fano resonances intorduced in coupled plasmon-quantum emitter systems \cite{Tasgin13,clark2000second}. By the use of carefully designed MNSs enhanced nonlinear conversion of light can be achieved. 

We construct a coupled MNS system composed of an AgNP of $\sim$100 nm size and an AgNW of $\sim$60 nm diameter and several $\mu$m length on a dielectric (glass) surface. When a 1064 nm near-infrared (NIR) excitation laser is focused solely on an isolated AgNW, a weak 532 nm SHG signal is generated. On the contrary, when the coupled system is excited in the vicinity of the coupling region, the observed SH signal increases up to 30 fold. We have demonstrated this enhancement by performing an {\it in situ} experiment in which, as the focus is moved along the AgNW axis, it is found that the generated SH signal increases by a factor of 30 when the focus passes through the AgNP coupled region. The SHG level at AgNW body a few $\mu$m away from the AgNP is at the same level as that of an isolated AgNW. The SH signal from a bare AgNP is found to be only 1/6 of the isolated AgNW. The observed enhancement can be obtained in our model by coupling the converter with a MNS which has a 10 times longer plasmon lifetime, see Fig.~2.

\section{Theoretical model}

In this section, we develop a theoretical model describing the second harmonic response of a system of two coupled plasmonic oscillators. We introduce the effective Hamiltonian for the system and derive the equations of motion for the plasmon polariton mode fields. We numerically time evolve the equations to obtain the steady state occupations of the plasmon modes and show that second harmonic conversion generation can be either enhanced or suppressed by choosing an appropriate MNS supporting a particular frequency mode.

We treat the MNSs as if they do not have a spatial extent, i.e. they are point particles. In the physical situation, however, the particles must have sizes and hence there must be retardation effects. In ref. \cite{Turkpence14}, coupling of a MNS and a quantum emitter object is treated with the same theoretical model. It is shown that the simple model can well predict the amount and the spectral position of the SHG enhancement by comparing the results with the simulations performed by using MNPBEM Toolbox in MATLAB \cite{Hohenester12}. The simulations take the retardation effects into account, the physical objects are simulated with their true geometries. 

\subsection{The model system}

\begin{figure}[h]
\includegraphics[width=3.2in]{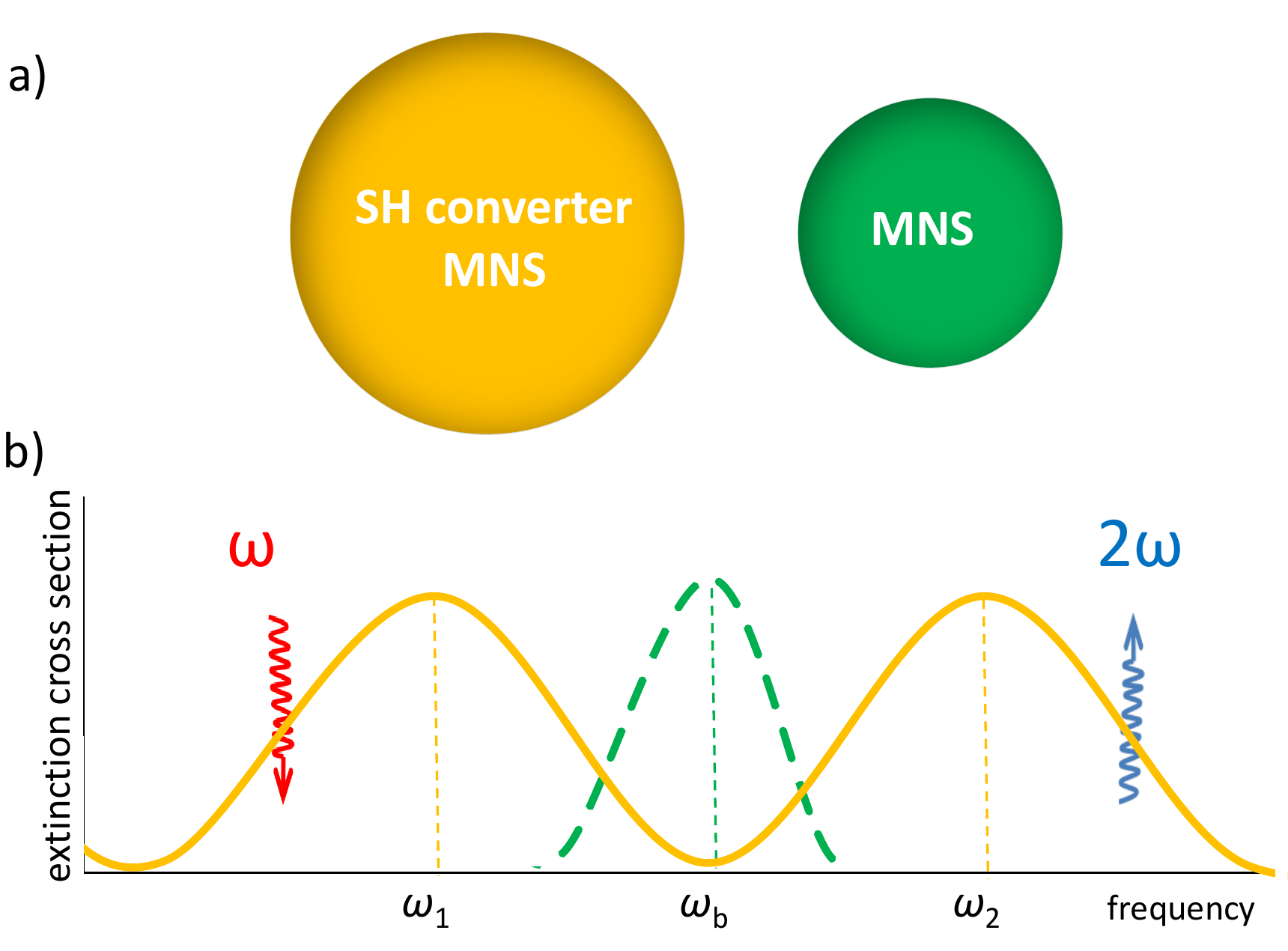}
\caption{(color online) A representative picture of the coupled oscillators, and their resonances. First oscillator (converter) supports two modes, $\omega_{1}$ and $\omega_{2}$, where the two frequencies are approximately equal to the driving and second harmonic field frequencies respectively. Second oscillator is coupled with the first oscillator introducing path interference effects in the second harmonic response.}
\label{fig:Figure1}
\end{figure}

We consider two MNSs which support plasmonic excitations at optical wavelengths. The MNSs interact with each other due to induced charge oscillations. The first MNS supports two plasmon modes of resonances $\omega_{1}$ and $\omega_{2}$ in the relevant frequency regime, see Fig.~\ref{fig:Figure1}b solid yellow curve. Shape of the second MNS is chosen so that it has a single plasmonic response peaked at $\omega_{b}$, see Fig. \ref{fig:Figure1}b dashed green curve, near the $\omega_2$ mode of the converter MNS.

The converter MNS has a non-centrosymmetric shape  (the spherical shape in Fig. \ref{fig:Figure1}a is for demonstrative purposes) which enables the SH process. In the model, we consider only the two plasmon modes in which the converter is driven ($\omega_1$) by the strong laser ($\omega$) and the mode ($\omega_2$) into which the conversion process ($2\omega$) takes place. In general, however, the converter MNS may support many plasmon modes in between $\omega_1$ and $\omega_2$ modes or out of the region covered by the two modes. We neglect the coupling of $\omega$ and $2\omega$ excitations to other modes for the sake of obtaining a simple model. We aim a basic physical picture. The $\omega_2$ mode of the converter is coupled to the $\omega_b$ mode of the second MNS. Only for being able to track the leak of the $2\omega$ oscillations from $\omega_b$ mode to other modes in the converter MNS, we also take the coupling of $\omega_b$ with $\omega_1$ mode into account.

\subsection{Hamiltonian}

Dynamics of the system is as follows. The incident strong driving field,  $\varepsilon_p e^{-i\omega t}$, is coupled to the $\omega_1$ mode of the converter and induces plasmon oscillations with frequency $\omega$. The plasmon field ($\hat{a}_1$) produced by the localized surface plasmon excitation yields to a strong electromagnetic field in the plasmonic converter. This localized strong field enables multiphoton (plasmon) processes come into play. The field oscillating at frequency $\omega$, trapped in the $\omega_1$ plasmon polarization field, gives rise to the second harmonic polarization oscillations at $2\omega$ in the $\omega_2$ mode. Two plasmons in the $\omega_1$ mode combines \cite{finazzi2012plasmon} and generates a $2\omega$ plasmon in the $\omega_2$ mode. It is experimentally demonstrated \cite{grosse2012nonlinear} that SHG process takes place in a plasmonic materials through plasmons. Conversion is carried through such a mechanism since the overlap integral of the process, $\chi^{(2)}\sim E_1^2({\bf r})E_2^*({\bf r})$, highly increases due to the localization of both plasmon modes \cite{ginzburg2012nonlinearly}. The $\omega_b$ mode of the second MNS interacts with both $\omega_{1}$ and $\omega_{2}$ modes. The interaction strengths can be tuned either by varying $\omega_b$ or by changing the position or shape of the second MNS (i.e. changing the overlap integral).   

The total Hamiltonian can be written as the sum of the energies of the plasmons in the first ($\hat{H}_a$) and the second ($\hat{H}_b$) MNSs, the interaction between the MNSs ($\hat{H}_{int}$), the energy supplied by the incident field ($\hat{H}_p$), and the term governing the second harmonic generation process ($\hat{H}_{sh}$),

\begin{equation}
\hat{H}=\hat{H}_a+\hat{H}_b+\hat{H}_{int}+\hat{H}_p+\hat{H}_{sh}. \label{H}
\end{equation} 
\noindent
where

\begin{eqnarray}
\hat{H}_a=\hbar\omega_1\hat{a}_1^\dagger\hat{a}_1+\hbar\omega_2\hat{a}_2^\dagger\hat{a}_2, \label{H_a} \\
\hat{H}_b=\hbar\omega_b\hat{b}^\dagger\hat{b}, \label{H_b} \\
\hat{H}_{int}=\hbar(f_1\hat{a}_1^\dagger\hat{b}+f_1^*\hat{a}_1\hat{b}^\dagger)+\hbar(f_2\hat{a}_2^\dagger\hat{b}+f_2^*\hat{a}_2\hat{b}^\dagger), \label{H_int} \\
\hat{H}_{p}=i\hbar(\hat{a}_1^\dagger\epsilon_p e^{-i\omega t}-\hat{a}_1\epsilon_p^* e^{i\omega t}), \label{H_p} \\
\hat{H}_{sh}=\hbar\chi^{(2)}(\hat{a}_2^\dagger\hat{a}_1\hat{a}_1+\hat{a}_1^\dagger\hat{a}_1^\dagger\hat{a}_2). \label{H_sh}
\end{eqnarray}

\noindent
$\hat{a}_1$ and $\hat{a}_2$ are the annihilation operators for the collective plasmon excitations in the first MNS, corresponding to the modes with resonance frequencies $\omega_1$ and $\omega_2$ respectively. Similarly, $\hat{b}$ is the annihilation operator for $\omega_b$ mode of the second MNS. ($\hat{a}_1$, $\hat{a}_2$ and $\hat{b}$ will represent the amplitudes of the related plasmon oscillations.) $f_1$ ($f_2$) is the coupling matrix element between the field induced by $\hat{a}_1$ ($\hat{a}_2$) mode of the SH converter (first MNS) and the $\hat{b}$ mode of the second MNS. Eq. (\ref{H_p}) describes the energy contribution of the incident field which drives $\hat{a}_1$ PP mode. In Eq. (\ref{H_sh}), two low energetic plasmons in the $\hat{a}_1$ mode oscillating at $\omega$, combine to generate a second harmonic plasmon in the $\hat{a}_2$ mode, oscillating at $2\omega$. The parameter $\chi^{(2)}$, in units of frequency, is proportional to the second harmonic susceptibility of the oscillator.   

We use Heisenberg equation of motion, $i\hbar\dot{\hat{A}}=[\hat{A},\hat{H}]$ for any operator $\hat{A}$, to derive the time evolution of the plasmon amplitudes. After obtaining the quantum dynamics of each mode, since we are not interested in the entanglement features but only in intensities, we substitute the quantum operators $\hat{a}_1$, $\hat{a}_2$, and $\hat{b}$ with their expectation values, $\alpha_1$, $\alpha_2$, and $\alpha_b$. We plug in the damping rates for the PP fields, namely $\gamma_1$, $\gamma_2$, and $\gamma_b$  respectively. Consequently we obtain the following equations for the plasmon amplitudes.

\begin{equation}
\dot{\alpha}_1=(-i\omega_1 -\gamma_1)\alpha_1-f_1\alpha_b+\epsilon_pe^{-i\omega t}-2i\chi^{(2)}\alpha_1^*\alpha_2 \label{alpha_1}
\end{equation}
\begin{equation}
\dot{\alpha}_2=(-i\omega_2 -\gamma_2)\alpha_2-f_2\alpha_b+i\chi^{(2)}\alpha_1^2 \label{alpha_2}
\end{equation}
\begin{equation}
\dot{\alpha}_b=(-i\omega_b -\gamma_b)\alpha_b-f_1\alpha_1-if_2\alpha_2 \label{alpha_b}
\end{equation}

In the steady state, the oscillation modes can only support the driving frequency ($\omega$) and the SH generated frequency ($2\omega$) of the form

\begin{eqnarray}
\alpha_1=\alpha_1^{(1)}e^{-i\omega t}+\alpha_1^{(2)}e^{-i2\omega t}, \label{mode_a1} \\
\alpha_2=\alpha_2^{(1)}e^{-i\omega t}+\alpha_2^{(2)}e^{-i2\omega t}, \label{mode_a2} \\
\alpha_b=\alpha_b^{(1)}e^{-i\omega t}+\alpha_b^{(2)}e^{-i2\omega t}. \label{mode_ab} 
\end{eqnarray}
\noindent
Amplitudes $\alpha_1^{(1)}$, $\alpha_2^{(1)}$ and $\alpha_b^{(1)}$ are the amplitudes of linear plasmon oscillations ($\omega$), and $\alpha_1^{(2)}$, $\alpha_2^{(2)}$ and $\alpha_b^{(2)}$ are the amplitudes of the second harmonic ($2\omega$) plasmon oscillations. 

We numerically time evolve equations (\ref{alpha_1},\ref{alpha_2},\ref{alpha_b}) and then obtain the time behaviour after steady state has been reached. Using the Fourier transform technique we determine the steady state amplitudes, $\alpha_1^{(2)}$, $\alpha_2^{(2)}$ and $\alpha_b^{(2)}$, the amplitudes of second harmonic oscillations. The number of SH plasmons can be determined by summing over the plasmons generated in all three modes as

\begin{equation}
N_{sh}\equiv\ {\vert\alpha_1^{(2)}\vert}^2+{\vert\alpha_2^{(2)}\vert}^2+{\vert\alpha_b^{(2)}\vert}^2. \label{no_of_sh_plasmons}
\end{equation}

\begin{figure}[h]
\includegraphics[width=3.2in]{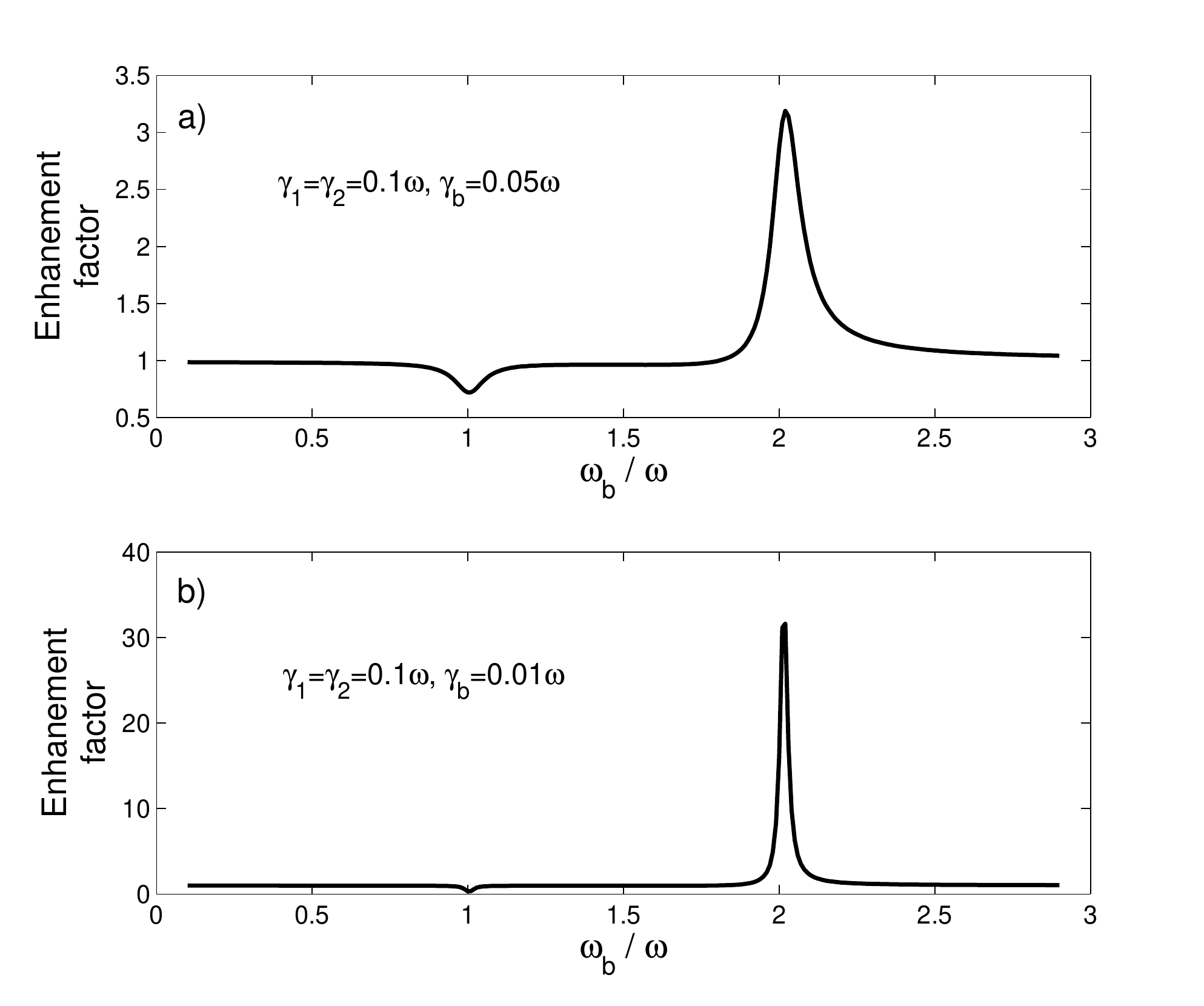}
\caption{Enhancement factor with changing resonance frequency of the second MNS, $\omega_b$ for the coupling of the MNSs with decay rates, a) $\gamma_1 = \gamma_2 = 0.1\omega$ and $\gamma_b = 0.05\omega$, b) $\gamma_1 = \gamma_2 = 0.1\omega$ and $\gamma_b = 0.01\omega$. In both cases $\epsilon_p=0.01\omega$.}
\label{fig:Figure2}
\end{figure}

We note that $e^{-i2\omega t}$ oscillations in $\alpha_1$ and $\alpha_b$ modes are not generated in these modes. They are rather transferred from $\alpha_2$ mode due to interactions.

\subsection{Tuning the second harmonic conversion}

We are interested in the enhancement and suppression of the SHG from the first MNS, in the presence of alternative absorption/emission pats due to the coupling with the second MNS. We compare the number of $2\omega$ plasmons generated in the presence of coupling ($f_1\neq0, f_2\neq0 $), $N_{sh}$, to the number of $2\omega$ plasmons in the absence of coupling ($f_1=0, f_2=0$), $N_{sh}^{(0)}$. Thus, we define the enhancement factor as the ratio of $N_{sh}$ to $N_{sh}^{(0)}$.
\begin{equation}
\mbox{Enhancement factor} \equiv \frac{N_{sh}}{N_{sh}^{(0)}}
\end{equation}

In order to identify the spectral positions of the nonlinear Fano resonances, we calculate the SHG enhancement factor for different resonance frequencies, $\omega_b$, of the second MNS. In our simulations, we fix the parameters $\omega_1=\omega$, $\omega_2=2\omega$, $\gamma_1 = \gamma_2 = 0.1\omega$, $f_1=0.2\omega$ and $f_1=0.8\omega$. In Fig.s \ref{fig:Figure2}a and \ref{fig:Figure2}b, we set the decay rate of the second MNS as $\gamma_b=0.05\omega$ and $\gamma_b=0.01\omega$, respectively.

Fig. \ref{fig:Figure2}a, where $\gamma_b=0.05\omega$, shows that in the case of two coupled MNS, SHG can be enhanced more than three times as compared to the case of the first MNS alone. Maximum enhancement is obtained when the resonance of the second MNS is tuned to $\omega_b=2.02\omega$. On the contrary, SHG can be suppressed to 0.72 when $\omega_b$ is tuned to $\omega_b=1\omega$. Fig. \ref{fig:Figure2}b shows that when a second MNS with smaller damping ($\gamma_b=0.01\omega$) is used, a smaller suppression but much larger and enhancement can be obtained. At $\omega_b=2.02\omega$ ($\omega_b=1\omega$), enhancement (suppression) factor of 32 (0.31) is found. 

Emergence of enhencement at about $\omega_b \approx 2\omega$ can be understood using the arguments given in ref. \cite{Turkpence14}. The off-frequency term, that is $\omega_2-2\omega$, is cancelled by an auxiliary term which emerges due to the path interference effects (see the discussion below Eq. (9)in ref. \cite{Turkpence14}). Suppression at $\omega_b=1\omega$ take place since the linear path interference does not allow the excitation at the driving frequency $\omega$. When the linear response is suppressed, smaller $\omega$ ($\hat{a}_1$) plasmon intensities result in less SH conversion.  	 	

In the following section, we report the results of an experiment on a silver nanowire-silver nanoparticle system locally illuminated with a Gaussian beam source.

\section{Experiment}

In this section we explain the details of the experimental study on a similar system to the one studied within the theoretical model, a system of two coupled plasmonic oscillators; a silver nanowire and a silver nanoparticle.

\subsection{Colloidal solution preparation procedure} 

Silver nanowires are synthesized by self-seeding polyol process. In this technique, an inorganic salt is reduced by a polyol and agglomeration of particles is prevented by addition of surfactant which is commonly polyvinylpyrrolidone (PVP). Required chemicals are bought from Sigma-Aldrich. 7 mg of NaCl is added into 10 mL of 0.45 M ethylene glycol (EG) solution of PVP and heated at $170 \celsius$. By using injection pump, solution of 0.12 M AgNO$_3$ in 5 mL of EG is added drop-wise at a rate of 5 mL/h. During this process, solution is stirred at 1000 rpm rate by magnetic stirrer. After the drop-wise EG addition process, solution is heated up at $170 \celsius$ for 30 minutes and is cooled to room temperature. To enable the removal of polymer from the solution, the diluted solution with acetone is centrifuged two times for 20 minutes at 6000 rpm. Afterwards, wires are dispersed in ethanol and centrifuged again under the same conditions. At the end of this procedure, nanowires with 60 nm diameter and length of 8 to 10 $\mu$m are obtained. 

\begin{figure}[h]
\includegraphics[width=3.2in]{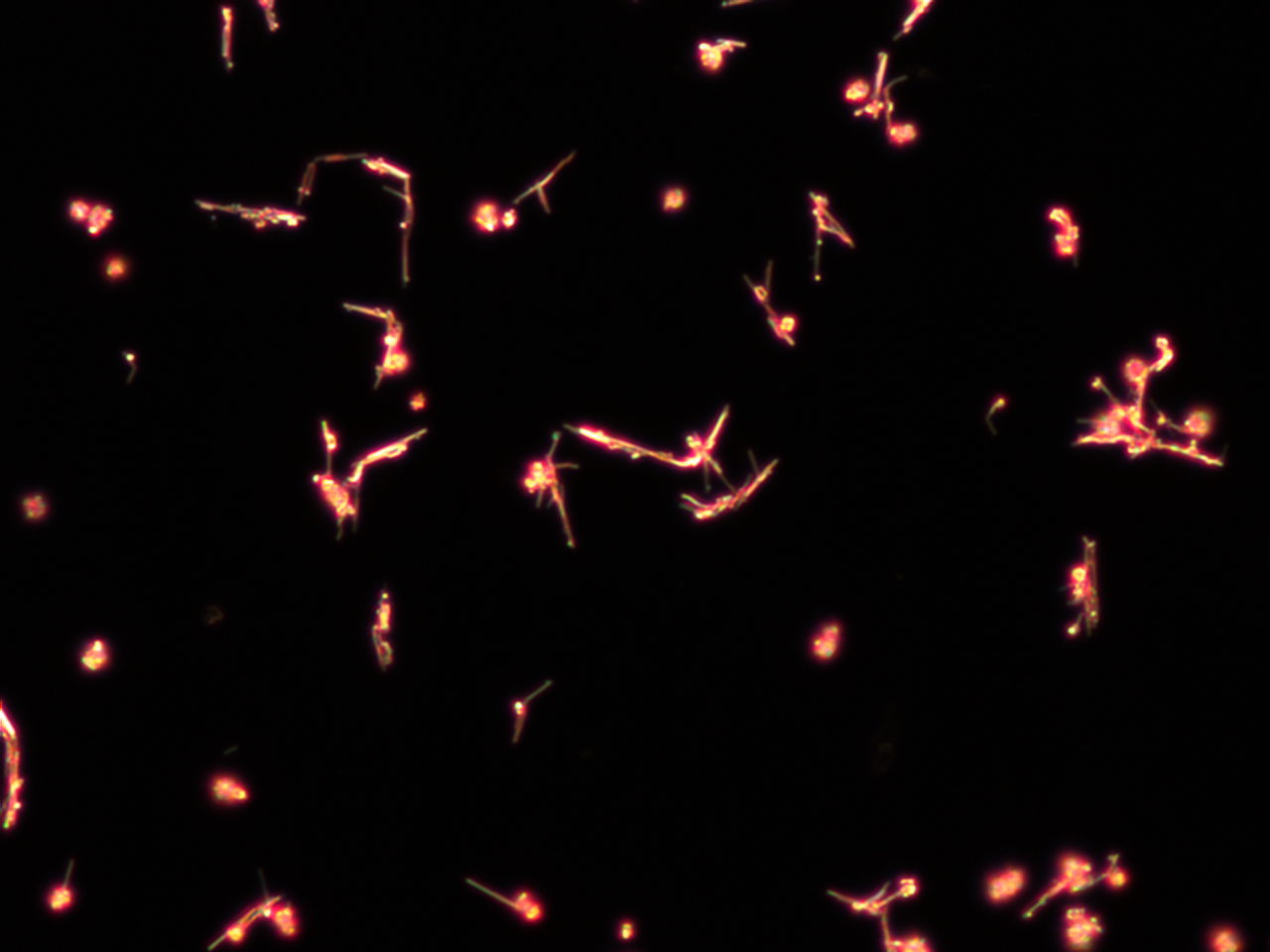}
\caption{A darkfield image of the AgNW - AgNP network deposited on microscope cover-slip. Light colored long straight features in the image represent the AgNWs and the bright intense spots represent the AgNPs.}
\label{fig:Figure3}
\end{figure}

Solution containing bipyramids of metallic silver is synthesized with polymer-mediated polyol process as well. 94 mM AgNO$_3$ is added to 3 mL EG solution. Another 3 mL of EG solution containing 0.11 mM NaBr and 144 mM PVP is prepared and two solutions are added dropwise into heated EG solution of 5 mL in an oil bath at 160${\celsius}$ which contains 30 $\mu$L of 10 mM NaBr. After 5 hours, bipyramid solution with an average size of 150 nm is obtained \cite{Coskun11,Wiley06}. The synthesis procedure leaves the AgNWs and AgNPs with a polymer coating of about 3-4 nm thickness.

The resulting AgNPs are co-deposited on the same substrates of AgNWs to obtain a network of  NW-NP complexes, which are ultimately asymmetric structures and cover a macroscopically large area. Such coated surfaces are first imaged by dark-field microscopy to ensure the existence of the complexes (Fig. \ref{fig:Figure3}). The AgNPs are observed to display different colors such as green and blue under white light illumination as a result of their size dependent plasmon resonance in the visible range. The AgNWs are observed to show a reddish glimmer as a result of shift of their plasmon resonance towards NIR due to high aspect ratio of their geometry. A combination of such two Ag nanostructures is what we expect to boost the SHG due to conversion of NIR plasmons to visible plasmons upon interaction of NWs with NPs.

\begin{figure}[h]
\includegraphics[width=3.2in]{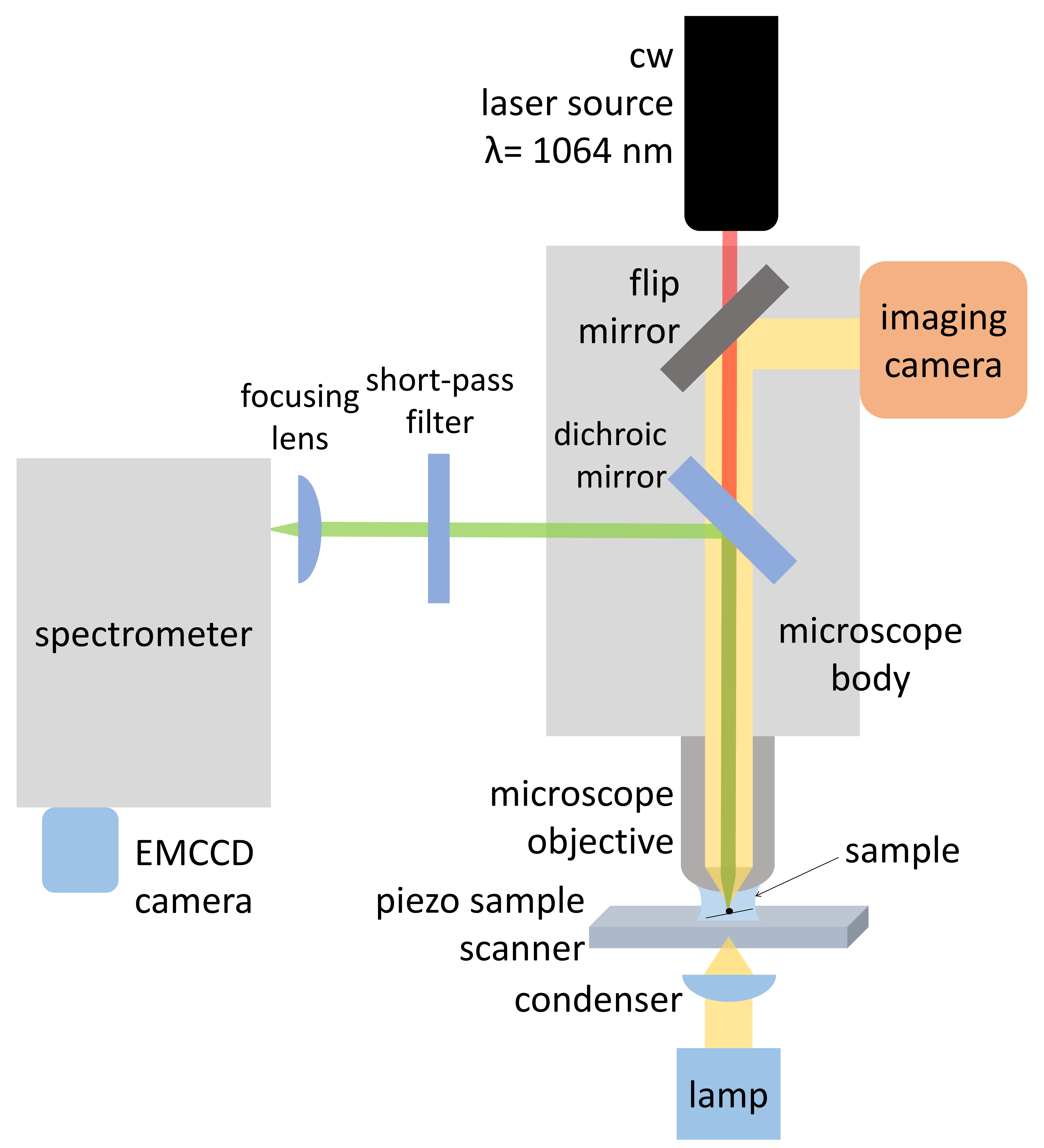}
\caption{The experimental setup is composed of a cw Nd:YAG laser with $\lambda = 1064$ nm, a long-pass dichroic mirror, a flip mirror, a microscope with objective, a piezo sample scanner, a short-pass filter (cut-off at 800 nm), a focusing lens, a spectrometer with EMCCD camera, an illumination lamp, a condenser, and an imaging camera.}
\label{fig:Figure4}
\end{figure}

\subsection{Optical measurements}

In Fig. 5, we show that the AgNWs do possess a manifold of plasmon resonances comparable to the sketched model in Fig. 1. The data shows differential absorption (Fig. 5b) which is calculated from the original absorption spectrum measurement (Fig. 5a) performed on a liquid dispersion of AgNWs at 0.125 mg/ml concentration. The dominant plasmon resonance \cite{kottmann2001plasmon,zhu2011high} centered at 390 nm (769 THz) is subtracted from the measured data by performing a peak fit. The residual differential absorption curve (in black) is displayed in THz units to allow for direct comparison to Fig 1. In Fig. 5 we also show three Gaussian peaks out of which two of them are much like the $\omega_{1}$, $\omega_{2}$ (orange solid) plasmon bands in the model.

\begin{figure}[h]
\includegraphics[width=3.2in]{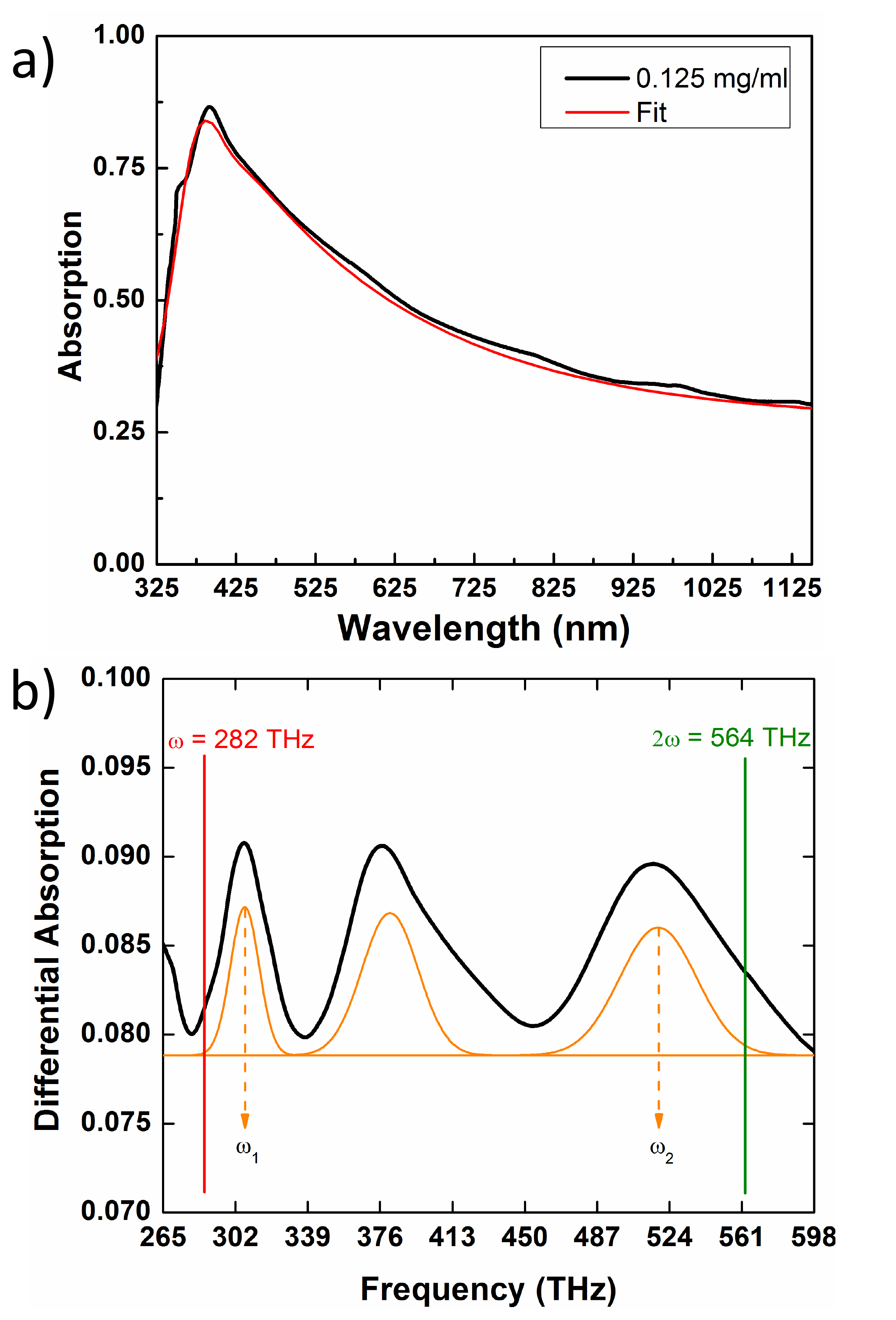}
\caption{a) The absorption spectrum of liquid dispersion of AgNW (black) and a peak fit to the strongest plasmon mode at 390 nm (red), b) The differential absorption curve after the contribution of the strong plasmon peak is subtracted (black), two Gaussian peaks representing the $\omega_1$, $\omega_2$ (orange solid) plasmon bands.}
\label{fig:Figure5}
\end{figure}

We used an inverted microscope (Zeiss model Axiovert 200) with a 63X (1.4 NA) objective lens for our experiments. In Fig. \ref{fig:Figure4}, a scheme of experimental setup is displayed. We used a cw Nd:YAG laser (Cobolt model Rumba) with wavelength of $\lambda = 1064$ nm and 500 mW output power as excitation source. The NIR signal was delivered from the back port of microscope, resulting in an intensity of 40 MW/cm$^2$ on the sample. Resulting SHG from the AgNW-AgNP cluster was registered by a spectrometer (Andor models Shamrock 750 spectrograph + Newton 971 EMCCD) in corporation with a long-pass dichroic mirror cutting off the backscattered 1064 nm signal and reflecting a band of 520-750 nm.

\begin{figure}[h]
\includegraphics[width=3.2in]{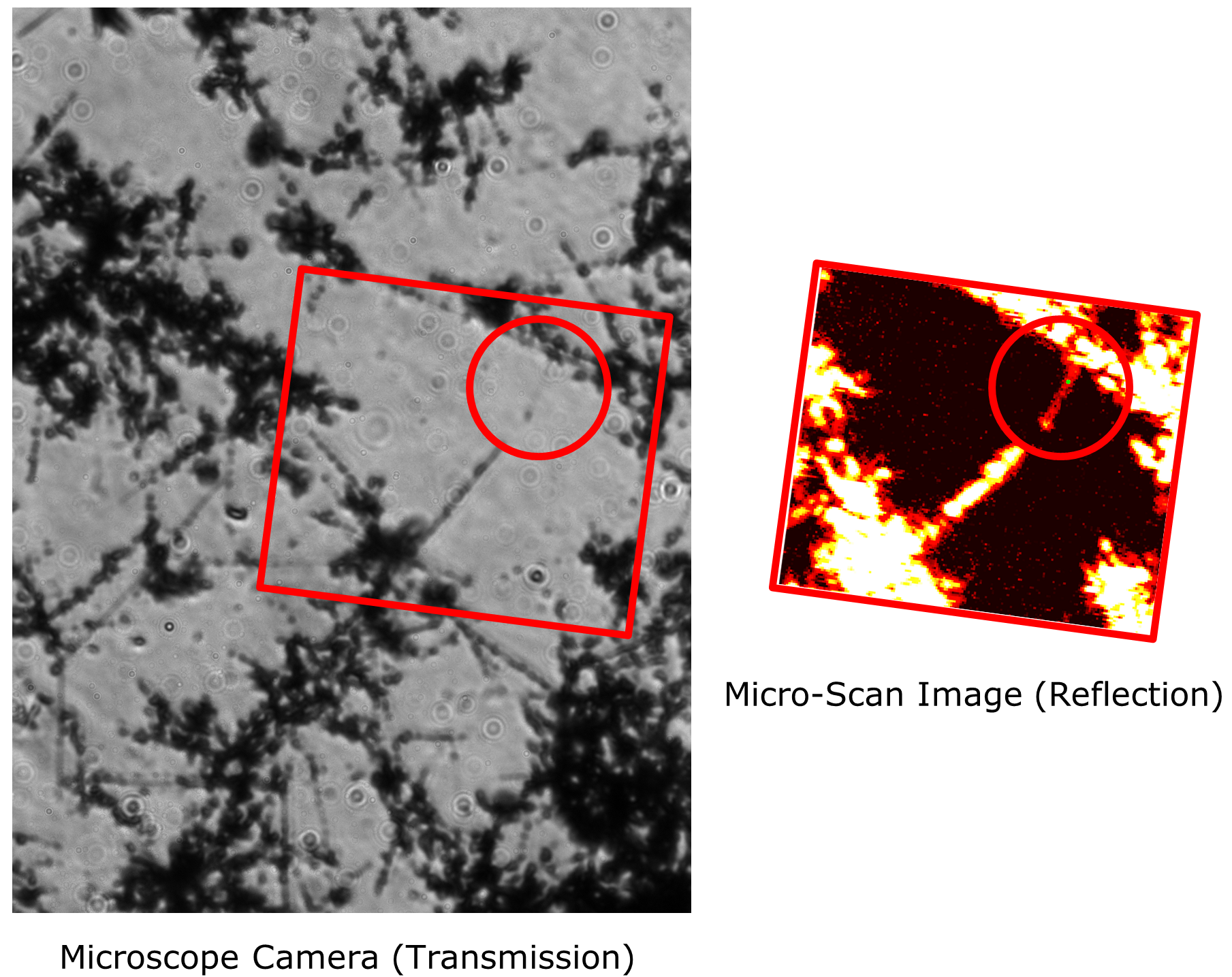}
\caption{A transmission image by the microscope camera on the left and a micro-scan reflection image of the area as indicated by the red rectangle, on the right. The red circles show a region with a single AgNW with a single AgNP attached close to its end.}
\label{fig:Figure6}
\end{figure}

We first record a transmission image by the visual inspection camera that works in the visible, see Fig. \ref{fig:Figure6}. The dark figures in the image indicate the location of AgNWs, AgNPs  and their clusters with a very poor resolution. Nevertheless this image is sufficient to locate a candidate region that bears a single AgNW-AgNP complex for further detailed study. The red circle on the transmission image on the left points a region of possible interest with one NW and NP complex. The red rectangle indicates the region which is scanned at a high resolution by the piezo-stage. The reflection image produced by the scanning of the same region is given on the right using a 1064 nm NIR laser. The reflection image produced at the illumination wavelength shows intense reflection from crowded clusters as well as a single straight feature which is the isolated part of a single AgNW and an AgNP attached towards the end of the AgNW. The faint straight shadow on the left image and a dark spot at its end appear as a straight bright feature on the left with a brighter spot at its end that indicates the attached AgNP. 

\begin{figure}[h]
\includegraphics[width=3.2in]{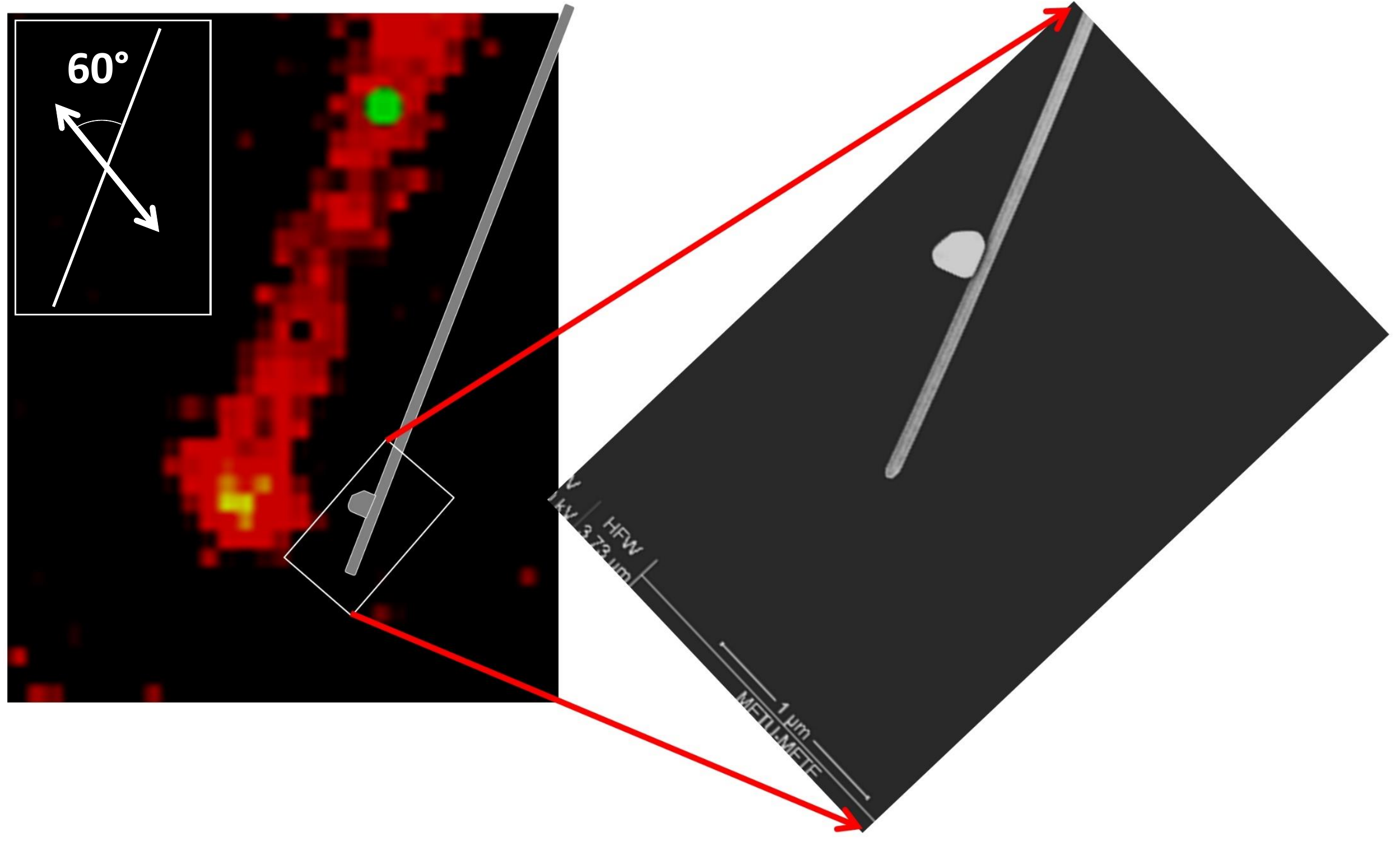}
\caption{A zoom in into the encircled region in the micro-scan image along with a sketch of AgNW-AgNP complex. The SEM image on the right shows the end of the AgNW with high magnification.}
\label{fig:Figure7}
\end{figure}

Such AgNW-AgNP complexes are often encountered upon inspection of these samples with scanning electron microscope. Fig. \ref{fig:Figure7} shows a zoomed version of the encircled straight part of the micro-scan image along with a representative sketch of the AgNW-AgNP complex drawn to scale. An SEM image is provided on the right that shows a highly magnified image of the end of AgNW with the attached AgNP along its body. The AgNW is typically of 50-60 nm in diameter and can be as long as 5-10 $\mu$m. The AgNPs are typically of 50-150 nm size. 

\subsection{Experimental results}

\begin{figure}[h]
\includegraphics[width=3.2in]{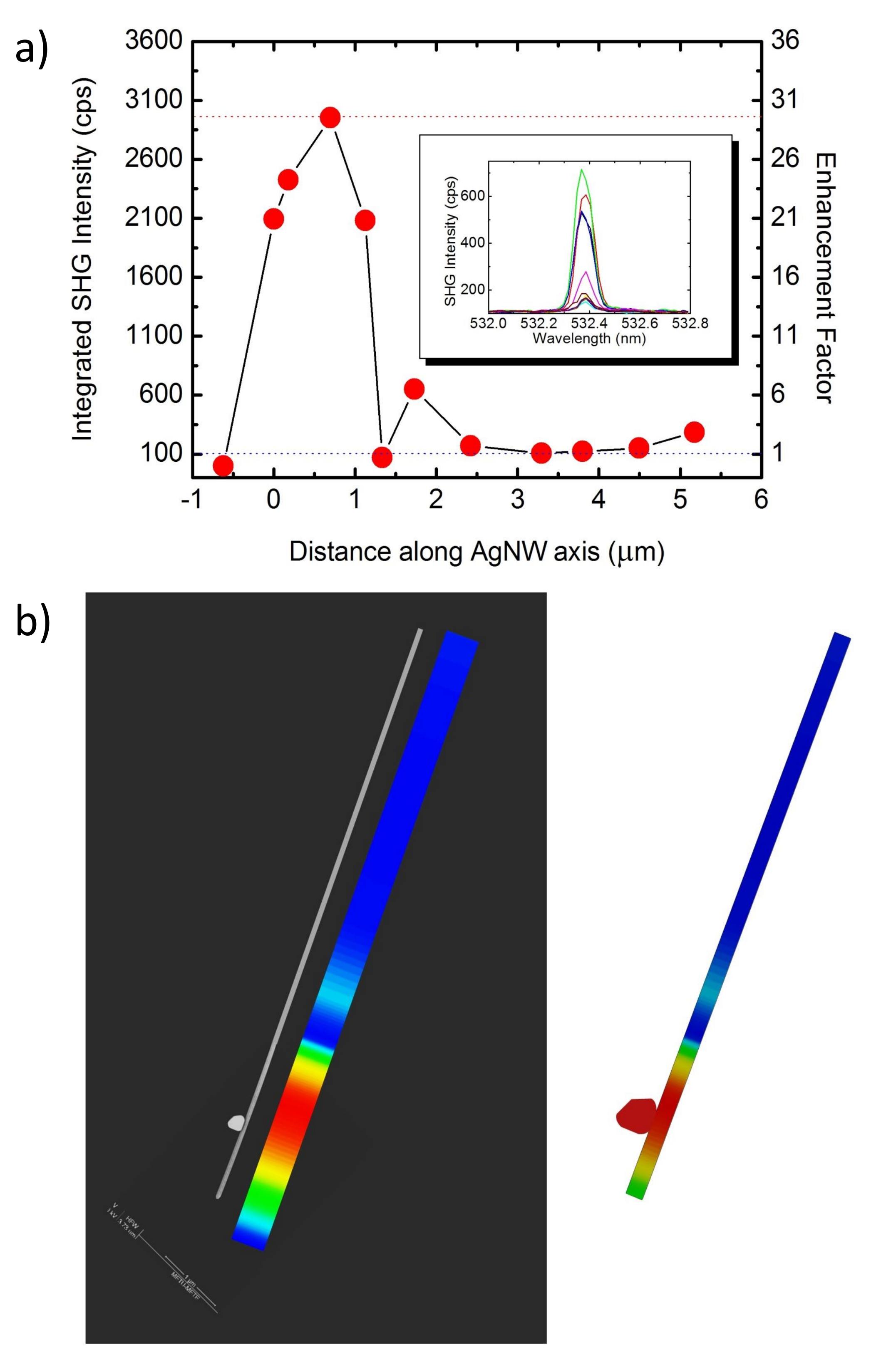}
\caption{a) The SHG spectra obtained at different positions along the body of the AgNW (inset). The integrated SHG intensity (left vertical axis) and the enhancement factor (right vertical axis) as a function of position on the AgNW. The breadth of the SHG signal peak along the AgNW axis is due to the 1 $\mu$m focal spot size. b) The SHG integral signal intensity as a color bar and the AgNW along with it, clearly show that the enhancement originates from coupling of the AgNW with the AgNP. The color bar superposed onto the representative sketch of the system - on the right.}
\label{fig:Figure8}
\end{figure}

By the help of the micro-scan image, we can direct our focal spot at any pixel of choice and acquire SHG spectra using a spectrometer  EMCCD combination. While moving the diffraction limited illumination spot stepwise along the axis of the AgNW, a SHG spectrum is acquired at each step and the acquired spectra are plotted (Fig. \ref{fig:Figure8}a inset). These spectra are acquired for 60 s and integrated over SHG band. The integral SHG signal is plotted against the distance along the AgNW axis (Fig. \ref{fig:Figure8}a). The excitation polarization is linear and is in the plane of the sample surface. The white arrow in the left inset in Fig.~\ref{fig:Figure7} shows the polarization direction which is 60$^o$ to this particular AgNW that we studied.

There are 3 major results that we obtain. (i) When the illumination spot is not located on the AgNW, no SHG signal is registered. So the SHG originates genuinely from the plasmonic structure. (ii) We observe SHG even with cw illumination (50 mW on the sample) from AgNW plasmonic structures. (iii) The SHG signal is enhanced by a factor of about 30 when the illumination spot is on the AgNP-AgNW complex with respect to the body of the AgNW alone.

A better visual represantation of the SHG intensity distribution as a function of position along the AgNW axis is given in Fig. \ref{fig:Figure8}b. A color bar is produced from the integral SHG signal and displayed in parallel with the actual SEM image and its representative straight extension. The SHG signal clearly appears starting from the end of the AgNW and reaches its maximum around where the AgNP is attached to the AgNW.  An overlap of the color bar with a represantation of such an AgNW-AgNP plasmonic hybrid complex on the right clearly depicts the observed effect: An enhanced SH conversion spot is constructed by the hybridization of the AgNW and AgNP plasmons.

\section{Discussion and conclusions}

In this paper we study the tunability of SHG from a coupled system of metal nano particle and a nanowire. Our experiments show that SHG from nanoparticles or nanowires can be enhanced about 30 times as compared to the uncoupled ones, because of path interference effects. We also introduce a single theoretical model and demonstrate the origin of the enhancement in our experiment. 

Silver nanoparticle and nanowire samples are illuminated with continuous wave NIR laser source of 1064 nm wavelength at different focal points with a focal spot size 1 $\mu$m diameter. First, AgNW and AgNP samples of typical sizes 5 $\mu$m and 100 nm respectively, are illuminated separately and SH response is recorded. Second, the two MNSs are combined and a representative coupled MNS is identified. The focus of NIR laser is moved stepwise along the axis and the SH response at 532 nm wavelength is recorded  at each step (Fig. \ref{fig:Figure8}). It is observed that when the laser focus is in the region containing the coupled AgNP, the SH signal is enhanced up to 30 times as compared to the AgNP-free regions on the AgNW axis.

Our experimental results can be interpreted by the help of the developed theoretical model examining the effects of coupling of a MNS with a SH converter MNS. We show that the enhancement factor ($\sim$30) observed in the experiment can be obtained by attaching a higher quality MNS to the SH converter MNS. The attached MNS has a distorted bipyramidal shape and can support dark modes \cite{panaro2014dark} with relatively longer lifetimes than that of bright modes. Therefore in our experiment AgNW plays the role of SH converter and bipyramid AgNP is the coupled higher quality plasmonic oscillator. This is also confirmed by control experiments performed on AgNW and AgNP samples separately, where the former is found to generate $\sim$6 times higher SH conversion. It can be noted that the AgNP size effect is incorporated to our model through the $f_{1}$ and $f_{2}$ parameters in Eqn. \ref{H_int}. AgNPs with different sizes are expected to have different $\omega_{b}$ positions which result in different levels of overlaps between $\omega_{1}$ and $\omega_{2}$ and hence different magnitude of $f_{1}$ and $f_{2}$, respectively. The magnitudes of these parameters play an important role in the SHG enhancement.

Alzar et. al. show that absorption cancellations due to Fano resonances can be achieved in entirely classical systems \cite{Alzar02}, which are observed in our and recent experiments \cite{Renger11,Thyagarajan13,Berthelot12,Wunderlich13,luk2010fano}. We approach the problem with a simple model of two coupled oscillators which have different damping rates. We reveal the enhancement/suppression effects on the spectrum of the nonlinear conversion using simple arguments.

\section*{Acknowledgements}
A.B. acknowledges support from TUBITAK Grants No. 113F239, 113M931, 113F375, 114E105 and METU grant No. BAP-01-05-2015-003. M.E.T. acknowledges support from TUBITAK Grants No. 112T927 and No. 114F170. We would like to thank Dr. Marco Lazzarino and Dr. Denys Naumenko from Laboratorio TASC, CNR-IOM for useful discussions and help with the experiments. We also thank Dr. Oguz Gulseren from Bilkent University, Ankara, Turkey for his support.

\bibliography{mybibfile}

\end{document}